\documentclass{article}

\usepackage{microtype}
\usepackage{graphicx}
\usepackage{subfigure}
\usepackage{booktabs} % for professional tables
\usepackage{amsmath}

\usepackage{bm}
\usepackage{hyperref}

\usepackage[accepted]{icml2021}

\icmltitlerunning{Accounting for Selection Effects in Supernova Cosmology with Simulation-Based Inference and Hierarchical Bayesian Modelling}

\begin{document}

\twocolumn[
\icmltitle{Accounting for Selection Effects in Supernova Cosmology with Simulation-Based Inference and Hierarchical Bayesian Modelling}
%\icmltitle{Under review at ICML 2024 AI for Science workshop}

% It is OKAY to include author information, even for blind
% submissions: the style file will automatically remove it for you
% unless you've provided the [accepted] option to the icml2021
% package.

% List of affiliations: The first argument should be a (short)
% identifier you will use later to specify author affiliations
% Academic affiliations should list Department, University, City, Region, Country
% Industry affiliations should list Company, City, Region, Country

% You can specify symbols, otherwise they are numbered in order.
% Ideally, you should not use this facility. Affiliations will be numbered
% in order of appearance and this is the preferred way.

\begin{icmlauthorlist}
\icmlauthor{Benjamin M. Boyd}{to}
\icmlauthor{Matthew Grayling}{to}
\icmlauthor{Stephen Thorp}{sw}
\icmlauthor{Kaisey S. Mandel}{to}

\end{icmlauthorlist}
\icmlcorrespondingauthor{Benjamin M. Boyd}{bmb41@cam.ac.uk}
\icmlaffiliation{to}{University of Cambridge, United Kingdom}
\icmlaffiliation{sw}{Stockholm University, Sweden}

% You may provide any keywords that you
% find helpful for describing your paper; these are used to populate
% the "keywords" metadata in the PDF but will not be shown in the document
\icmlkeywords{Machine Learning, ICML}

\vskip 0.3in
]

% this must go after the closing bracket ] following \twocolumn[ ...

% This command actually creates the footnote in the first column
% listing the affiliations and the copyright notice.
% The command takes one argument, which is text to display at the start of the footnote.
% The \icmlEqualContribution command is standard text for equal contribution.
% Remove it (just {}) if you do not need this facility.

%\printAffiliationsAndNotice{}  % leave blank if no need to mention equal contribution
\printAffiliationsAndNotice{} % otherwise use the standard text.

\begin{abstract}
%This document provides a basic paper template and submission guidelines.
%Abstracts must be a single paragraph, ideally between 4--6 sentences long.
%Gross violations will trigger corrections at the camera-ready phase.

Type Ia supernovae (SNe Ia) are thermonuclear exploding stars that can be used to put constraints on the nature of our universe. One challenge with population analyses of SNe Ia is Malmquist bias, where we preferentially observe the brighter SNe due to limitations of our telescopes. If untreated, this bias can propagate through to our posteriors on cosmological parameters. In this paper, we develop a novel technique of using a normalising flow to learn the non-analytical likelihood of observing a SN Ia for a given survey from simulations, that is independent of any 
cosmological model. The learnt likelihood is then used in a hierarchical Bayesian model with Hamiltonian Monte Carlo sampling to put constraints on different sets of cosmological parameters conditioned on the observed data. We verify this technique on toy model simulations finding excellent agreement with analytically-derived posteriors to within $1 \sigma$. 

\end{abstract}
\section{Introduction}
\label{sec:intro}

Type Ia supernovae (SNe Ia) are violent explosions of white dwarf stars when they interact with their binary partner. All SNe Ia share a similar peak brightness and can be used as ``standardisable candles''. The difference between the modelled absolute brightness and apparent brightness of a standard 
candle is
used in astronomy to estimate distance. The redshifts  of SNe, caused by the universe's expansion, can be plotted against the distances estimates in a Hubble diagram to put constraints on cosmology.

In standardisation, it has been observed that SNe Ia are brighter if they are bluer, and also if they evolve more slowly around peak brightness. These effects are conventionally incorporated into a linear model \cite{tripp1998} describing the relationship between SNe Ia absolute brightnesses $M_s$ (in magnitudes where lower is brighter) and observable parameters: 
\begin{equation}\label{eq:tripp}
M_s = m_s - \mu(z_s;\,\bm{C}) = M_0 + \alpha \times x_s + \beta \times c_s + \epsilon 
\end{equation}
where $m_s$ represents the peak apparent $B$ band magnitude of SN $s$. The distance modulus $\mu$ is a function of the cosmological redshift $z_s$ and cosmology $\bm{C}$. The other observable SNe parameters include the peak $B-V$ apparent colour $c_s$ (where lower is bluer and higher is redder) and stretch parameter $x_s$,  encoding how long the light curve takes to decay (where higher is longer). The remaining scatter $\epsilon$ has residual variance $\sigma_\text{res}^2$. The three parameters $\bm{d_s}=(m_s,c_s,x_s)^T$ can be defined and fitted using the SALT light curve model \citep{guy2007,guy2010}, widely adopted in SNe Ia analysis. The effect of the stretch and colour on the absolute magnitude is parametrized by global hyperparameters $\alpha$ and $\beta$. These hyperparameters, the absolute magnitude constant $M_0$ and cosmology $\bm{C}$ can be inferred by jointly analysing the overall SN sample and their observed redshifts $\hat{z}_s$. We denote estimated or measured values as ``hatted'' variables, 
which differ from the true, latent values by some random measurement error.

When conducting any type of cosmological inference using SNe Ia, it is important to consider selection effects, otherwise they can propagate to biases in cosmological parameter inference. A common type of bias is known as Malmquist bias  \citep{malmquist1922}, where surveys are more likely to detect bright objects. If this is true for SNe Ia surveys, then it would also create a bias in detecting bluer and longer lasting SNe as they are brighter.

The leading method to handle this bias is BEAMS with Bias Corrections \citetext{BBC, \citealt{kessler2017}} that uses state-of-the-art survey simulations from \textit{SNANA} \citep{kessler2009}. One-dimensional BBC involves splitting simulated SNe Ia into redshift bins. Each simulated SN Ia is fit by a model to estimate its distance modulus. For the SNe within each bin, the differences between fitted distance moduli and true distance moduli are calculated and averaged to make $\Delta \mu_{\text{bias}}(\hat{z}_s)$. To apply bias corrections on real SNe Ia, the learnt $\Delta \mu_{\text{bias}}(\hat{z}_s)$ is added to every distance modulus in the same bin. After this, the corrected Hubble diagram may be plotted and cosmology inferred. BBC may also be done in multiple dimensions, for example SNe Ia can also be be split into bins according to colour $\hat{c}_s$ and stretch $\hat{x}_s$. The process works similarly, where the bias in distance for each bin is averaged and future real SNe Ia are corrected accordingly. Importantly, the true distance moduli used in this method rely on an assumed cosmology, which we would like to avoid in future analysis. Furthermore, this method performs analysis in separate stages where the data is modified using BBC before the cosmological model is fit. 

Another approach to SN Ia cosmological analysis involves modelling the observable parameters in Equation \eqref{eq:tripp} within a hierarchical Bayesian model (HBM) \citep{march2011,shariff2016}. Malmquist bias can be accounted for by assuming the selection function $P(I_s=1\big|\,\bm{\hat{d}_s})$ is analytical, such as a step function or Gaussian cumulative distribution function. We indicate a SN has been detected with $I_s=1$. The selection function can be incorporated in a HBM framework \cite{rubin2015,march2018} where the data likelihood $P(\bm{\hat{d}_s}|\,\bm{\Theta})$ can be analytically modified to down-weight the probability of detecting a dim SN and up-weight the probability of detecting a bright one, such that:
\begin{equation}
\label{eq:HM}
P\big(\bm{\Theta} \big|\, \bm{\hat{D}}\big)  \propto \bigg[\prod_{s=1}^{N_{\text{SN}}} P\big(\bm{\hat{d}_s}\big|\,I_s=1,\hat{z}_s,\bm{\Theta}\big) \bigg]P\big(\bm{\Theta}\big)
\end{equation}
\begin{multline}
\label{eq:genlik}
P(\bm{\hat{d}_s}|\,I_s=1,\hat{z}_s,\bm{\Theta}) = \\\frac{P(I_s=1\big|\,\bm{\hat{d}_s})P(\bm{\hat{d}_s}|\,\hat{z}_s,\bm{\Theta})}{\int^\infty_{-\infty}P(I_s=1\big|\,\bm{\hat{d}_s})P(\bm{\hat{d}_s}|\,\hat{z}_s,\bm{\Theta})\,\text{d}\bm{\hat{d}_s}}
\end{multline}

where $\bm{\hat{D}}=\{\bm{\hat{d}_s},\hat{z}_s,I_s=1\}_{s=1}^{N_{\text{SN}}}$ and $\bm{\Theta}=(\bm{C},M_0,\alpha,\beta,...)$. These methodologies have their benefits since they are each one integrated analysis. While the assumption of a simple analytic ansatz for the selection function may be appropriate to first order, the true survey selection function induced by Malmquist bias is not known to be analytic.

Simulation-based inference (SBI) provides a way forward for the field by which the state-of-the-art simulations that capture the complex selection effects can be utilised and applied in cosmological analysis. SBI can be used to learn probability distributions and then apply them to real data to perform inference. This can be done using neural density estimation \cite{rezende2015,papamakarios2021} or neural ratio estimation \cite{hermans2020}. Previous examples of using 
SBI to address SN Ia selection effects in cosmological analysis have modelled the overall posterior conditioned on the entire SN sample at once $P\big(\bm{\bm{\Theta}} \big|\, \bm{\hat{D}}\hspace{0.1cm}\big)$ \citep{alsing2019,karchev2023,karchev2023b,karchev2024}. Since the current combined SN sample sizes are of the order of thousands, the current methods rely on data compression to learn the complex relationship between the full SN sample and cosmology. Importantly these methods need to be trained on a wide range of cosmological parameter values for a given cosmological model, which is expensive in terms of computational cost. The non-hierarchical approaches are also sensitive to the size of the sample so that the SN analysis model is required to be retrained if a single SN is added or removed from the sample. 

In this paper, we propose a new method that combines SBI with hierarchical Bayesian modelling for SN Ia cosmological analysis. This enables a non-analytical auxiliary density estimator, approximating the three-dimensional likelihood, to be trained to learn the survey selection effect independent of cosmology. The same likelihood may then be re-used in  
HBMs to allow for constraints on different cosmological models. The HBM framework is flexible to allow  for varying sample-sizes without the need to retrain the likelihood.

\section{Method}
\label{sec:method}

In this section, we outline the method used for our hierarchical simulation-based inference model used in our SN Ia cosmological analysis. In Section \ref{sec:bhm} we define the HBM, while in Section \ref{sec:nf} we describe the normalising flow used in likelihood estimation and finally in Section \ref{sec:toy} we outline the analytical toy simulations used to verify the method. 

\subsection{Hierarchical Bayesian Model}
In this section, we define the HBM \citetext{\citealt{mandel2017} Appendix D} used in our cosmological analysis. The  posterior takes the same form as Equation \eqref{eq:HM}, where in our model $\bm{\Theta} =(\bm{C}, M_0,\alpha, \beta,\sigma_{\text{res}}^2,c_0, \alpha_c,\sigma_c^2,x_0,\sigma_x^2)$. The population hyperparameters are described as follows:
\begin{itemize} 
\item $\bm{C}$: vector containing cosmological parameters to be fit for a given cosmological model. For example, in a flat $w$CDM model we might fit $w$ and $\Omega_{m0}$ or for a $\Lambda$CDM model we might fit for $\Omega_{m0}$ and  $\Omega_{\Lambda0}$.

\item $M_0$: the absolute magnitude constant shown in Equation \eqref{eq:tripp} is
the expected absolute magnitude for a SN
with light curve shape $x_s=0$ and colour $c_s=0$.

\item $\alpha$: the slope of the trend of absolute magnitude vs. light curve shape from Equation \eqref{eq:tripp}.
\item  $\beta$: the slope of the trend of absolute
magnitude vs. colour from Equation \eqref{eq:tripp}.
\item $\sigma_{\text{res}}^2$: the remaining residual variance around the mean trend
of absolute magnitude vs. light curve
shape and colour.
\item $c_0$: the expected colour for a SN Ia with
light curve shape $x_s$ = 0. If $\alpha_c=0$ then $c_0$ is
the population mean colour.
\item $\alpha_c$: the slope of the trend of colour vs.
light curve shape. 
\item $\sigma_{c}^2$: the variance around the mean trend
of colour vs. light curve shape.
\item $x_0$: the mean of the $x_s$ light curve shape population
distribution.
\item $\sigma_{x}^2$: the variance of the $x_s$ light curve shape population distribution.
\end{itemize}
The latent peak apparent magnitude $m_s$ is modelled using Equation \eqref{eq:tripp}. The additional scatter is drawn from $\epsilon \sim N(0,\sigma_{\text{res}}^2)$. The latent peak $B-V$ colour $c_s$ is modelled using:
\begin{equation}
    c_s = c_0 + \alpha_c x_s + \epsilon_c
\end{equation}
The additional colour scatter is drawn from $\epsilon_c \sim N(0,\sigma_{c}^2)$. The latent stretch $x_s$ is modelled using:
\begin{equation}
x_s  = x_0 + \epsilon_x
\end{equation}
the additional stretch scatter is drawn from  $\epsilon_x \sim N(0,\sigma_x^2)$. 

Observed redshifts $\hat{z}_s$ are perturbed away from the cosmological redshifts $z_s$ due to peculiar velocities, such that
 $\hat{z}_s=(1+z_s)/(1+z_{\text{pec}})-1+\epsilon_z$ \cite{carr2022}.
Peculiar redshifts are sampled from  $z_{\text{pec}}\sim N(0,\sigma_{\text{pec}}^2/c^2)$ where $\sigma_{\text{pec}}$ is the peculiar velocity dispersion and $c$ is the speed of light. Additional scatter is drawn from $\epsilon_z \sim N(0,\hat{\sigma}_{z,s}^2)$ where $\hat{\sigma}_{z,s}$ is the observed redshift measurement uncertainty. 

In inference, we only have access to the observed redshift $\hat{z}_s$ rather than the latent cosmological redshift $z_s$. To derive the HBM likelihood, we must employ an analytic approximation to incorporate this redshift error such that
$
P(\mu_s|\,\hat{z}_s,\bm{C}) \approx N(\mu_s|\,\mu(\hat{z}_s;\,\bm{C}),\sigma_{\mu|\,z,s}^2)$
where the variance is approximated as $\sigma_{\mu|\,z,s}^2\approx\big(\sigma_{\text{pec}}^2/c^2+\hat{\sigma}_{z,s}^2\big)\big|\partial\mu(z_s;\boldsymbol{C})/\partial z_s\big|_{z_s=\hat{z}_s}^2
$ \cite{mandel2009,march2011}.

The likelihood of the latent parameters $\bm{d_s}=(m_s,c_s,x_s)^T$  given the data $\bm{\hat{d}_s}$ is modelled as a multivariate Gaussian such that
$P(\bm{\hat{d}_s}|\,\bm{d_s})=N(\bm{\hat{d}_s}|\,\bm{d_s}, \bm{W_s})$
where $\bm{W_s}$ is the error covariance matrix from the SALT light curve fitter \cite{guy2007,guy2010}. In our method, we implicitly marginalise over latent variables $\bm{d_s}$ as the approximated auxiliary density is only a function of hyperparameters and expected apparent magnitude.

\begin{figure}[] 
\vspace{-0.3cm}
\centering    
\includegraphics[width=0.34\textwidth]{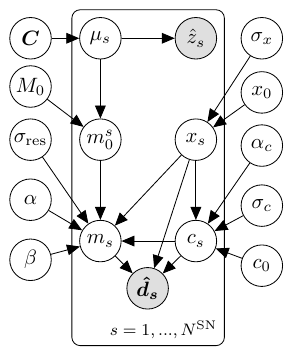}
\vspace{-0.3cm}
\caption[]{Graphical model illustrating the dependencies between hyperparameters, latent parameters and the observed data (shaded in grey) in the hierarchical Bayesian model.}
\label{fig:graphical_model}
\vspace{-0.4cm}
\end{figure}

The graphical model in Figure \ref{fig:graphical_model} illustrates the relationship between the above hyperparameters, latent and observed variables in our hierarchical model. We define the hyperpriors in Table \ref{priortab}. For sampling the hierarchical model we use Hamiltonian Monte Carlo \cite{duane1987}, a technique that uses gradient information (something that neural network architectures also use) for faster convergence on the posterior. We use the No-U-Turn Sampler (NUTS) in \textit{numpyro} \citep{pyro, numpyro}, a probabilistic programming package for automatic differentiation and just-in-time (JIT) compilation built with \textit{JAX} \cite{jax2018github}.
\label{sec:bhm}
\subsection{Training the Normalising Flow to Approximate the Likelihood}
\label{sec:nf}
\begin{figure*}[] 
\vspace{-0.25cm}
\centering    
\includegraphics[width=0.7\textwidth]{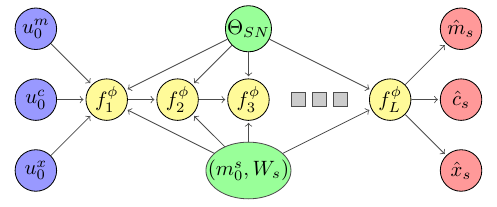}
\vspace{-0.5cm}
\caption[]{Illustration of how parameters interact with the normalising flow that models $q_\phi \big(\bm{\hat{d}_s}\big|m^s_0,\bm{W_s},\bm{\Theta_{\text{SN}}}\big)$ to approximate the auxiliary density $g$ for the observed data likelihood. The three $u_0$ parameters are all drawn from unit normal distributions and passed through $L$ transforms $f_l$ with learnable parameters $\phi$. The transforms are conditioned on the expected apparent magnitude $m_0^s$ for a given supernova (at zero stretch and colour), measurement covariances $\bm{W_s}$ and supernova hyperparameters $\bm{\Theta_{\text{SN}}} $. The outputs after $f_L$ are the three SALT parameters representing observed apparent magnitude $\hat{m}_s$, colour $\hat{c}_s$ and stretch $\hat{x}_s$}
\label{fig:flow_arch}
\vspace{-0.6cm}
\end{figure*}
For the observed data likelihood we write:
\begin{multline}
P\big(\bm{\hat{d}_s}\big|\,I_s=1,\hat{z}_s,\bm{\Theta}\big)=\\g\big(\bm{\hat{d}_s}\big|m^s_0=M_0+\mu(\hat{z}_s;\bm{C}),\bm{W_s},\bm{\Theta_{\text{SN}}}\big)
\end{multline}
 where $\bm{\Theta_{\text{SN}}}=(\alpha, \beta,\sigma_{\text{res}}^2,c_0, \alpha_c,\sigma_c^2,x_0,\sigma_x^2)$ contains the SN hyperparameters and $m_0^s$ is the expected apparent magnitude for a SN with zero colour and stretch such that $m_0^s=\text{E}[m_s|c_s=0,x_s=0,\hat{z}_s,M_0,\bm{C}]$. It is useful to define the auxiliary density $g\big(\bm{\hat{d}}\big|m_0,\bm{W},\bm{\Theta_{\text{SN}}}\big)$, to take advantage of the fact that the observed data likelihood depends on $(\hat{z}_s,M_0,\bm{C})$ only through the combination $m^s_0 = M_0 + \mu(\hat{z}_s;\bm{C})$.  This same auxiliary density can be reused to define the likelihood for constraining different cosmological models. 
 
 The auxiliary density can be learnt with simulation-based inference as we describe in Section \ref{sec:nf} or for some special cases derive it analytically as we do in Section \ref{sec:toy}.

\begin{table}[t]
\caption{Hyperparameters and respective hyperpriors used in our hierarchical  Bayesian model. Detailed also are the ranges used in the likelihood training. $U(B_L,B_U)$ is a uniform distribution between $B_L$ and $B_U$, while $HN(0,1^2)$ represents a positive half-normal distribution centred at 0 with a variance of 1.}
\label{priortab}
\begin{center}
\begin{small}
\begin{sc}
\begin{tabular}{lcccr}
\toprule
Hyperparameter & Hyperprior & Training Range\\
\midrule

$w_0$ (Flat $w$CDM)    &  $U(-2,2)$ & - \\
$\Omega_{m0}$ (Flat $w$CDM) & $U(0,1)$& -\\
$w_0$ ($\Lambda$CDM)   &  $U(0,2)$ & -  \\
$\Omega_{\Lambda 0}$  ($\Lambda$CDM)   &  $U(0,2)$  & -  \\  
$M_0$ & $U(-\infty,\infty)$ & -\\
$\alpha$ & $U(-\infty,\infty)$ & $U(-0.55,0.25)$\\
$\beta$ & $U(-\infty,\infty)$ & $U(0,4)$\\
$\sigma_{res}$ & $HN(0,1^2)$& $U(0,0.2)$\\
$c_0$ & $U(-\infty,\infty)$& $U(-0.4,0.4)$\\
$\alpha_c$ & $U(-\infty,\infty)$& $U(-0.03,0.03)$\\
$\sigma_{c}$ & $HN(0,1^2)$& $U(0,0.1)$ \\
$x_0$ & $U(-\infty,\infty)$ &$U(-1,1)$\\
$\sigma_{x}$ & $HN(0,1^2)$ & $U(0,2)$\\

\bottomrule
\end{tabular}
\end{sc}
\end{small}
\end{center}
\vskip -0.3in
\end{table}
To learn our three-dimensional auxiliary density $g \big(\bm{\hat{d}}\big|\, m_0,\bm{W},\bm{\Theta_{\text{SN}}}\big)$ we use a normalising flow \cite{tabak2010density,tabak2013family}. We select a normalising flow instead of a mixture density network \cite{bishop1994} since we require the likelihood is flexible enough to capture the non-analytical selection effects from realistic survey simulations. The normalising flow $q_\phi$, used to approximate $g$, is defined such that:
\begin{equation}
\label{eq:nf}
q_\phi \big(\bm{\hat{d}}\big|m_0,\bm{W},\bm{\Theta_{\text{SN}}}\big) = N(\bm{u_0} |\,\bm{0},\bm{I}) \prod^L_{l=1} \  \biggr\rvert \frac{\partial f_l^{\bm{\phi}}}{\partial \bm{u}_{l-1}}\biggr\rvert^{-1}
\end{equation}
where $\bm{u_0} \sim N(\bm{0},\bm{I})$, $\bm{u_l} = f_l^{\bm{\phi}}(\bm{u}_{l-1}|m_0,\bm{W},\bm{\Theta_{\text{SN}}})$  and $\bm{\hat{d}} = f^{\bm{\phi}}_L$.
The weights $\bm{\phi}$ are optimised by running stochastic gradient descent on batches of 2000 random simulations of individual SNe from the forward model using different input ($m^{\text{sim}}_0$, $\bm{W}_{\text{sim}}$, $\bm{\Theta}^{\text{sim}}_\text{SN}$) values. Each simulation comprises an individual observed supernova represented by a data-parameter pair ($\hat{\bm{d}}_{\text{sim}}$; $m^{\text{sim}}_0$, $\bm{W}_{\text{sim}}$, $\bm{\Theta}^{\text{sim}}_\text{SN})$.  At each iteration, the log of $q_{\phi}$, averaged over the batch of 2000 SNe, is used to compute the gradient. This optimisation is equivalent to minimising a Monte Carlo approximation of the forward Kullback-Leibler (KL) divergence between $g$ and $q_\phi$ \cite{papamakarios2021}. It is simple to evaluate normalising flows with the chain rule as their architectures are designed to have lower triangular Jacobian matrices, meaning the determinant is the product of diagonal terms. Figure \ref{fig:flow_arch} illustrates how the parameters are given to the normalising flow. For this work, we use the Masked Autoregressive Flow (MAF) architecture \cite{papamakarios2017masked} that involves stacking Masked Autoencoder for Distribution Estimation (MADE) blocks \cite{germain2015made}.

To train the normalising flow, each individual SN simulated is forward modelled from a different set of hyperparameters  $\bm{\Theta}^{\text{sim}}_{\text{SN}}$ that are sampled from a wide non-informative prior. We also sample $m^{\text{sim}}_0$ to ensure the normalising flow is not influenced by any cosmology $\bm{C}$. Furthermore it is necessary to put an emphasis on sampling the dimmest $m^{\text{sim}}_0$ values closer to the detection limit where Malmquist bias has the greatest effect. After these parameters are drawn, the SNe may be forward simulated according to Equation \eqref{eq:tripp} and the further equations in Section \ref{sec:bhm}. The survey simulation then determines which SNe are then detected. For state-of-the-art survey simulations such as those in \textit{SNANA} \cite{kessler2009}, this includes multiple complex factors such as telescope pointing plans, systematics and the weather. For the purpose of our tests the selection is determined by an analytical function of the observed data $\bm{\hat{d}_s}$ defined in Equation \eqref{eq:cdf}. It is only the data and parameters of the ``observed'' simulated SN ($I_{\text{sim}}=1$) that are given to the normalising flow for training $q_\phi \big(\bm{\hat{d}}\big|m_0,\bm{W},\bm{\Theta_{\text{SN}}}\big)$. Further details on how we train the normalising flow for our toy problem, including architectures and runtime, are detailed at the end of Section \ref{sec:toy}.

\subsection{Analytical Toy Simulations}
\label{sec:toy}

In this section, we define toy Malmquist bias SN simulations that have analytical likelihoods to validate the method outlined above. 

The 10,000 SNe that we validate the method on are simulated from the cosmological redshift distribution ${P(z_s)\propto(z_s+1)^{3/2}}$ in the range $0.01<z_s<2$. We convert these into distance modulus $\mu(z_s;\,\bm{C})$ using \textit{JAX-cosmo} \cite{jaxcosmo} conditioned on our cosmology $\bm{C}$. We define our true cosmology as a flat $\Lambda$CDM cosmology ($h=0.7324$, $\Omega_{m0}=0.28$, $\Omega_{\Lambda 0}=0.72$, $w_0=-1$ and 
 $w_a=0$). We later infer posteriors on a flat $w$CDM universe where $\bm{C_1}=(w_0,\Omega_{m0})$ and a potentially un-flat $\Lambda$CDM universe where 
 $\bm{C_2}=(\Omega_{m0},\Omega_{\Lambda 0})$. Our simulations are designed to replicate high-precision spectroscopic redshifts so we set $\hat{\sigma}_{z,s}=0$. The peculiar velocity dispersion $\sigma_{\text{pec}}$ is set to 200 kms$^{-1}$ and is used to forward model the observed redshifts $\hat{z}_s$.

After we have the distance modulus $\mu_{s}$ for each SN, we forward simulate the observed parameters $\bm{\hat{d}_s}$ from the same set of hyperparameters $\bm{\Theta}_{\text{SN}}$  using Equation \eqref{eq:tripp} and the equations detailed at the beginning of Section \ref{sec:bhm}. The true values of each hyperparameter used in the simulation are shown in Table \ref{results}. We assume a diagonal measurement covariance matrix $\bm{W}_s$. The measurement variances for apparent magnitudes are sampled according to $\ln W^{mm}_s \sim N(0.2(m_s-56),1.2^2)$, where $m_s$ is the apparent magnitude before the measurement error is applied. For apparent colour measurement variances we sample from $\ln W^{cc}_s \sim N(-7.0,0.6^2)$ and for apparent stretch measurement variances we sample from $\ln W^{xx}_s \sim N(-3,1^2)$.

\begin{figure}[] 
\vspace{-0.8cm}
\centering    
\includegraphics[width=0.48\textwidth]{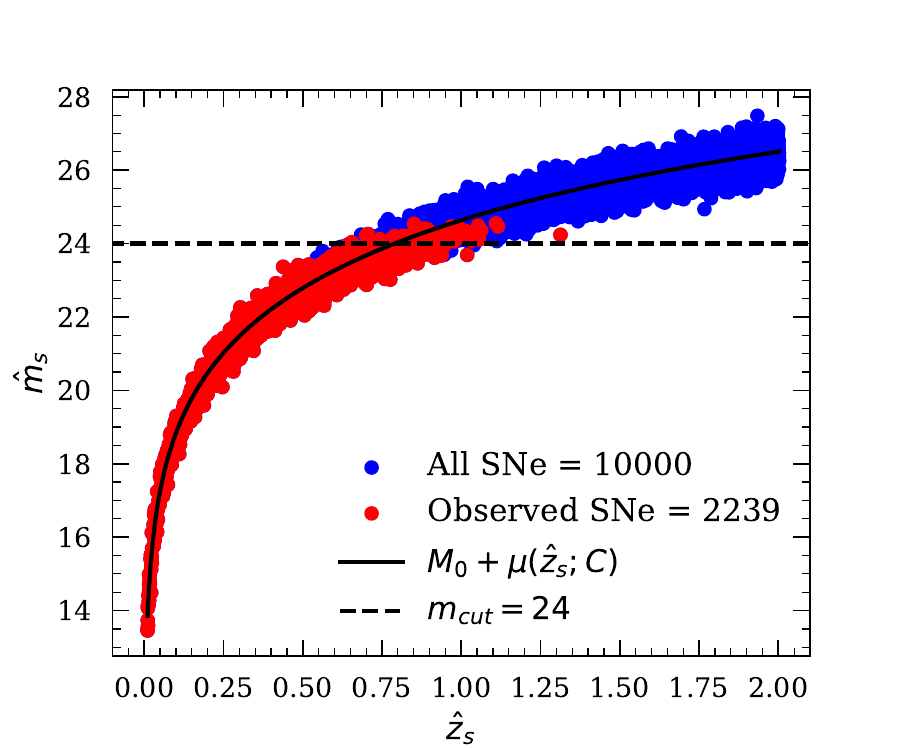}
\vspace{-0.8cm}
\caption[]{Toy simulations showing the effects of using the Gaussian cumulative distribution selection function to model Malmquist bias on the Hubble diagram. Plotted on the horizontal axis are the simulated spectroscopic redshifts and on the vertical axis are simulated apparent magnitudes of the SNe. In blue are all the SNe Ia simulated according to a flat $\Lambda$CDM universe ($h=0.7324,\Omega_{m_0}=0.28,\Omega_{\Lambda_0}=0.72,w_0=-1$). Over-plotted in red are the SNe that would have been detected according to a Gaussian cumulative distribution selection function centred at $m_{\text{cut}}=24$ with standard deviation $\sigma_{\text{cut}}=0.25$. The black solid line shows the effect of the true underlying cosmology that needs to be recovered. }
\label{fig:toy_hd}
\vspace{-0.6cm}
\end{figure}

\begin{figure*}[ht]

\centering    

\includegraphics[width=1\textwidth]{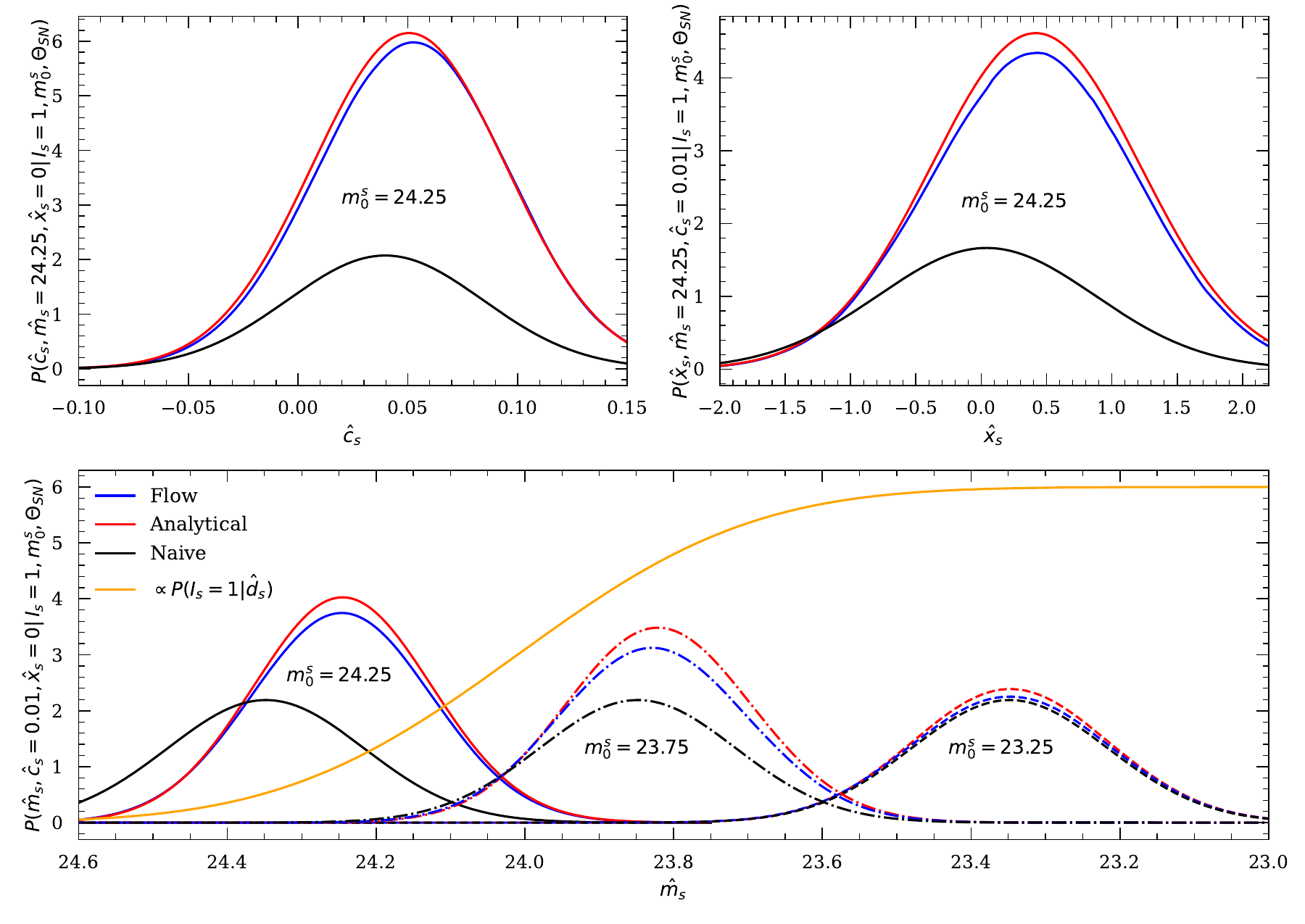}
\vspace{-1.2cm}
\caption[]{This plot compares the learnt normalising flow likelihood (blue) with the analytical data likelihood (red). We also plot the naive likelihood when it is just assumed to be Gaussian without considering selection effects in black. The bottom panel illustrates the likelihood as a function of apparent brightnesses for three different SNe. The selection function (orange) has a greater effect on dimmer SNe to the left of the plot. The top two panels illustrate the likelihood of the dimmest SN as function of observed colour and stretch. The other parameters used to generate these plots were set arbitrarily as $\bm{\Theta_{\text{SN}}}=(c_0 = 0.1 , x_0=-0.5 , \sigma_{\text{res}}= 0.1 , \sigma_c= 0.05 , \sigma_x = 1, \alpha =-0.1, \beta=2.7 , \alpha_c=0)$, $\sigma_{\mu|\,z,s}=0$, $W^{mm}_s=0.042$, $W^{cc}_s=0.031$ and $W^{xx}_s=0.25$.}
\vspace{-0.5cm}
\label{fig:like}
\end{figure*}
Following the simulation of SN observables $\bm{\hat{d}_s}$ we pass them through an analytical selection function to model Malmquist bias \cite{rubin2015} such that:
\begin{equation}
\label{eq:cdf}
P(I_s = 1 |\, \bm{\hat{d}_s}) = \Phi \Bigg(\frac{m_{\text{cut}} -(\hat{m}_s + a_{\text{cut}} \, \hat{x}_s + b_{\text{cut}} \, \hat{c}_s)}{\sigma_{\text{cut}}}\Bigg)
\end{equation} 
where $\Phi(\cdot)$ is a unit normal cumulative distribution function. $m_{\text{cut}}$ represents the limiting magnitude of the survey and $\sigma_{\text{cut}}$ represents the steepness of the drop-off around the limiting magnitude. The coefficient $a_{\text{cut}}$ is included to allow the possibility that longer lasting SNe are more likely to be detected and $b_{\text{cut}}$ allows the colour of the SN to influence the selection probability if the survey was not done in a $B$ band filter. For the purpose of four simulations we set $m_{\text{cut}}=24$, $\sigma_{\text{cut}}=0.25$, $a_{\text{cut}}=-0.1$ and $b_{\text{cut}}=-1$. After the selection function is defined, we perform one Bernoulli trial for each SN with their respective selection probability to determine if they are selected. In our subsequent analysis we then only use the 2,239 SN that have been selected after the Bernoulli trial ($I_s=1$). Figure \ref{fig:toy_hd} shows how the Malmquist bias selection function affects the SN that are observed.

\begin{table*}[t]
\vspace{-0.4cm}
\caption{In this table we present the mean and the standard deviation of the posteriors from our hierarchical models when applied to the toy simulations. We compare the true parameter values with the inferred normalising flow and analytical model posterior estimates.}
\label{results}

\begin{center}
\begin{small}
\begin{sc}
\begin{tabular}{lcccccc}
\toprule
Test &  &  \multicolumn{2}{c}{Flat $w$CDM} &  \multicolumn{2}{c}{$\Lambda$CDM}\\
\midrule

Hyperparameter & Truth  & Flow&Analytical & Flow &Analytical \\
\midrule
$w_0$&-1&-0.95$\pm$0.1&-0.94$\pm$0.11&fixed&fixed\\
$\Omega_{m0}$&0.28&0.26$\pm$0.04&0.25$\pm$0.05&0.26$\pm$0.04&0.26$\pm$0.04\\
$\Omega_{\Lambda 0}$&0.72&fixed&fixed&0.69$\pm$0.07&0.69$\pm$0.07\\
$M_0$&-19.392&-19.397$\pm$0.011&-19.393$\pm$0.012&-19.399$\pm$0.011&-19.394$\pm$0.012\\
$\alpha$&-0.154&-0.15$\pm$0.003&-0.152$\pm$0.003&-0.15$\pm$0.003&-0.152$\pm$0.003\\
$\beta$&2.252&2.28$\pm$0.05&2.24$\pm$0.05&2.28$\pm$0.05&2.24$\pm$0.05\\
$\sigma_{res}$&0.1&0.098$\pm$0.003&0.099$\pm$0.003&0.098$\pm$0.003&0.098$\pm$0.003\\
$c_0$&-0.061&-0.063$\pm$0.002&-0.06$\pm$0.002&-0.062$\pm$0.002&-0.06$\pm$0.002\\
$\alpha_c$&-0.008&-0.006$\pm$0.001&-0.007$\pm$0.001&-0.006$\pm$0.001&-0.006$\pm$0.001\\
$\sigma_c$&0.065&0.064$\pm$0.001&0.064$\pm$0.001&0.063$\pm$0.001&0.064$\pm$0.001\\
$x_0$&-0.432&-0.408$\pm$0.026&-0.407$\pm$0.026&-0.407$\pm$0.026&-0.407$\pm$0.026\\
$\sigma_x$&1.124&1.136$\pm$0.02&1.106$\pm$0.018&1.138$\pm$0.019&1.106$\pm$0.018\\
\bottomrule
\end{tabular}
\end{sc}
\end{small}
\end{center}
\vskip -0.3in
\end{table*}

\begin{figure*}[ht]
\vspace{-0.2cm}
\centering    

\includegraphics[width=1.\textwidth]{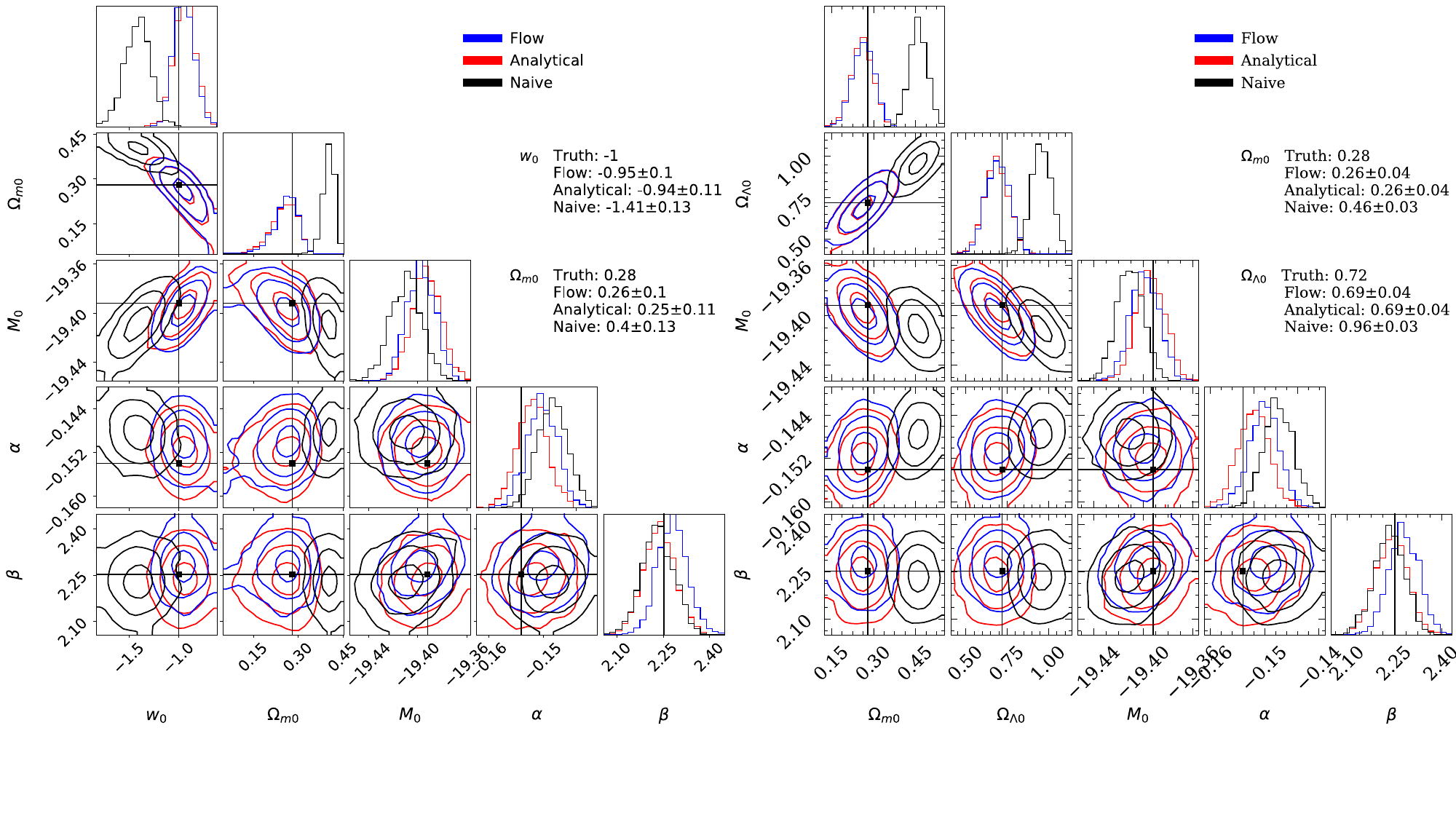}
\vspace{-2cm}
\caption[]{Corner plots showing inferred posteriors for our toy simulations. The left contours infer constrains on a flat $w$CDM cosmology whilst the right contours infer constraints on a $\Lambda$CDM cosmology. The other three parameters inferred are from Equation \eqref{eq:tripp} and describe the relationship between SN characteristics and absolute brightness. The blue contours are from the HBM that uses the normalising flow likelihood whilst the red contours use the analytical likelihood. We also include black contours that use a naive Gaussian likelihood that does not take selection effects into account. The true hyperparameters used in the simulations are indicated with black straight lines. }
\label{fig:corners}
\vspace{-0.5cm}
\end{figure*}

The auxiliary density for the observed SNe in the simulations defined above can be derived analytically. This  can done similarly to Equation \eqref{eq:genlik} such that:
\begin{multline}\label{eq:genliknew}
g \big(\bm{\hat{d}_s}\big|m^s_0,\bm{W_s},\bm{\Theta_{\text{SN}}}\big) = \\\frac{P(I_s=1\big|\,\bm{\hat{d}_s})P(\bm{\hat{d}_s}|\,m_0^s,\bm{\Theta_{\text{SN}}})}{\int^\infty_{-\infty}P(I_s=1\big|\,\bm{\hat{d}_s})P(\bm{\hat{d}_s}|\,m_0^s,\bm{\Theta_{\text{SN}}})\,\text{d}\bm{\hat{d}_s}}
\\= \frac{P(I_s=1\big|\,\bm{\hat{d}_s})N(\bm{\hat{d}_{s}}|\,\bm{\Psi_{d}},\bm{\Sigma_d})}{\int^\infty_{-\infty}P(I_s=1\big|\,\bm{\hat{d}_s})N(\bm{\hat{d}_{s}}|\,\bm{\Psi_{d}},\bm{\Sigma_d})\,\text{d}\bm{\hat{d}_s}}
\end{multline}
$N(\bm{\hat{d}_{s}}|\,\bm{\Psi_{d}},\bm{\Sigma_d})$
is a multivariate Gaussian with:

\[
\bm{\Psi_{d}} = 
\begin{pmatrix}
\text{E}[\hat{m}_s] \\
\text{E}[\hat{c}_s] \\
\text{E}[\hat{x}_s]
\end{pmatrix}
=
\begin{pmatrix}
m_0^s + \alpha x_0 + \beta (\alpha_c x + c_0) \\
\alpha_c x + c_0 \\
x_0
\end{pmatrix}
\]
\[
\bm{\Sigma_d} = \begin{pmatrix}
\text{Var}[\hat{m}_s] & \text{Cov}[\hat{m}_s,\hat{c}_s] &\text{Cov}[\hat{m}_s,\hat{x}_s]  \\
 \text{Cov}[\hat{m}_s,\hat{c}_s] & \text{Var}[\hat{c}_s]  &  \text{Cov}[\hat{c}_s,\hat{x}_s] \\
\text{Cov}[\hat{m}_s,\hat{x}_s]  & \text{Cov}[\hat{c}_s,\hat{x}_s]  & \text{Var}[\hat{x}_s]
\end{pmatrix}
\]
\begin{align*}
    \text{Var}[\hat{m}_s] &= \sigma_{\text{res}}^2+\sigma_{\mu|\,z,s}^2+\beta^2\sigma_c^2 \\
    &\hspace{1cm}+(\alpha+\beta\alpha_c)^2 \sigma_x^2+W^{mm}_s\\
    \text{Var}[\hat{c}_s] &= \sigma_c^2 + \alpha_c^2\sigma_x^2+W^{cc}_s\\
    \text{Var}[\hat{x}_s] &=\sigma_x^2+W^{xx}_s\\
    \text{Cov}[\hat{m}_s,\hat{c}_s] &= \alpha\alpha_c\sigma_x^2 +\beta(\sigma_c^2+\alpha_c^2\sigma_x^2)+W_s^{mc}\\
    \text{Cov}[\hat{m}_s,\hat{x}_s] &= \alpha\sigma_x^2 + \beta \alpha_c \sigma_x^2+W_s^{mx}\\
    \text{Cov}[\hat{c}_s,\hat{x}_s] &= \alpha_c \sigma_x^2 +W_s^{cx}
\end{align*}
The integral in the denominator of Equation \eqref{eq:genliknew} can be solved using the conditional properties of multivariate Gaussian distributions to split the likelihood into the product of three univariate Gaussian distributions. 
Appendix \ref{sec:proof} shows the three dimensions can be marginalised over one after another using the identity:
\begin{multline}
\label{eq:indentity}
\int^{\infty}_{-\infty} \Phi((\mu_1-x)/\sigma_1)N(x|\,\mu_2,\sigma_2^2)\,\text{d}x
 \\
= \Phi\Bigg(\frac{\mu_1 - \mu_2}{\sigma_1} \times \Big(1+\frac{\sigma_2^2}{\sigma_{1}^2}\Big)^{-\frac{1}{2}}\Bigg)
\end{multline}

We then proceed to train a normalising flow to learn from simulations the same likelihood we found analytically using the steps described in Section \ref{sec:nf}. The training set is constructed by forward modelling each individual SN $s$ 
 from a different set of hyperparameter values $\bm{\Theta}^{\text{sim}}_{\text{SN}}$. The training range for each hyperparameter is defined in Table \ref{priortab}. We sample 80\% of $m_0^{\text{sim}}$ from a negative half-normal distribution  $HN(m_{\text{cut}}+3\sigma_{\text{cut}},1^2)$ to ensure there are enough examples of dim objects at the detection limit where the selection effect has the greatest influence. The remaining 20\% of $m_0^{\text{sim}}$ examples are drawn from $U(11,m_{\text{cut}})$ to ensure the normalising flow has seen the full range of magnitudes.

 It helps in training to normalise the dimensions by first centering the data $\bm{\hat{d}_{\text{sim}}} \rightarrow \bm{\hat{d}_{\text{sim}}}-\bm{d_0^\text{sim}}$ where $\bm{d_0^\text{sim}}=(m_0^{\text{sim}},c^{\text{sim}}_0,x^{\text{sim}}_0)^T$. We also combine residual scatter  and observed redshift variances into a single conditional parameter such that $\sigma'_{\text{sim}}=\sqrt{\sigma_{\text{res}}^2+\sigma_{\mu|\,z,\text{sim}}^2}$. After this we do minimum/maximum normalisation across both data and parameters. In our example we stack $L=4$ MADE blocks, reversing the conditional order of the three dimensions in between each block. The architecture of the blocks includes two hidden layers of size 128 and ReLu activation functions. In total the normalising flow contains 143,384 learnable parameters. We use batches containing 2000 SNe and train on 250,000 random batches generated on-the-fly taking approximately 4 hours on a GPU. We implement the normalising flow using \textit{JAX} \cite{jax2018github} to allow for compatibility with the HBM. Sampling the HBM posterior that uses the normalising flow takes less than 20 CPU minutes with Hamiltonian Monte Carlo to produce 500 warm-up samples followed by 500 posterior samples for 4 chains.

After both the analytical likelihood is derived and the normalising flow likelihood is trained, we alternately use them in the hierarchical model defined in Section \ref{sec:bhm} to put constraints on different cosmological parameters and compare their results.

\section{Results and Discussion}

In this section, we present the results of our method on our toy simulations and discuss future work.

Before analysing the hierarchical model posteriors, we can take an intermediate step of validation by comparing the analytically derived likelihood and the normalising flow likelihood. Figure \ref{fig:like} compares the two for a given set of parameters. It is clear that the selection effect only comes into play for the dimmer SNe with magnitudes greater than $m_{\text{cut}}$. Brighter SN have likelihoods that are closer to a typical multivariate Gaussian. The normalising flow was able to account for selection effects in all three dimensions with a probability distribution functions close to the analytical solution. Despite having different centerings and normalisations, the likelihoods for the toy example look approximately Gaussian in shape. However, if $\sigma_{\text{cut}}$ decreases or more realistic simulations are used the shape of the likelihood becomes more skewed, hence why it is still important to use a flexible normalising flow. 

After the normalising flow was validated on the analytical likelihood solution, both were then used in hierarchical models to put constraints on cosmological parameters within two types of cosmological models as well as the other SN parameters. Table \ref{results} displays the means and standard deviations of the resulting posteriors as well as the true simulated values. Figure \ref{fig:corners} shows the inferred posteriors for the key hyperparameters. The posteriors utilising the normalising flow were able to recover the true simulated cosmology and supernova hyperparameters to within $2\sigma$ and agreed with the analytical posteriors within $1 \sigma$. The normalising flow likelihood was also flexible enough to be reused for both cosmological model fits reproducing almost identical posteriors on the SN hyperparameters. It is clear from these results that this technique combining SBI and hierarchical Bayesian modelling is effective in reproducing constraints on different cosmological parameters by correctly accounting for Malmquist bias.

Now that the method has been validated on simulations with analytical likelihoods, the next stage is to apply it to more realistic survey simulations with non-analytical likelihoods. The natural comparison is to train the normalising flow with the simulations from \textit{SNANA} \cite{kessler2009} and then compare inferred posteriors with the popular BBC method that uses binning. This work is already in progress and will provide useful insights to the effectiveness of BBCs uncertainty estimation and extent that it is independent of the fiducial simulated cosmology.

Other avenues for this work include hierarchical analysis on SNe from multiple different surveys at once. This is would be done by splitting the likelihood for $N_{\text{Surveys}}$ such that:
\begin{multline}
    P\big(\bm{\Theta} \big|\, \bm{\hat{D}}\big) \\ \propto \bigg[\prod_{j=1}^{N_{\text{Surveys}}}\prod_{s=1}^{N_{\text{SN}}^j} P_j\big(\bm{\hat{d}_s}\big|\,I_s=1,\hat{z}_s,\bm{\Theta}\big) \bigg]P\big(\bm{\Theta}\big)
\end{multline}
where $P_j\big(\bm{\hat{d}_s}\big|\,I_s=1,\hat{z}_s,\bm{\Theta}\big) $ is the observed data likelihood for survey $j$. Each survey has its own selection effect so this would require training $N_{\text{Survey}}$ normalising flows to model the individual likelihoods. SN Ia datasets that could be used for this type of analysis include the Pantheon+ \cite{scolnic2022} and Union \cite{rubin2023} compilations. 

Before conducting credible large scale cosmological analyses with this method, it is important to think about how survey systematics can be modelled. Survey systematics include correlations between SNe due to connecting factors, such as them being observed on the same evening or belonging to the same host galaxy. Traditionally, 
this information is captured by a $N_{\text{SN}} \times N_{\text{SN}}$ Gaussian covariance matrix, however, since we assume SNe are conditionally independent given SN hyperparameters, we currently do not include this. Explicitly conditioning on such latent common factors within our likelihoods could potentially solve this problem, but this requires further work.

Another use for this methodology would be to provide constraints on the astrophysics of SNe Ia. \textit{Simple-BayeSN} \cite{mandel2017} uses a similar hierarchical model framework but separates variations in intrinsic SN properties and extrinsic dust effects. This allows the model to make inferences on SN Ia dust environments including the ``mass-step'' \citep{kelly2010,sullivan2010, lampeit2010, childress2013} phenomenon where SNe Ia in larger host galaxies appear brighter. In principle our methodology is robust and flexible enough to be extended to these problems to allow for further analysis whilst still correcting for Malmquist bias.

\section{Conclusion}

We have developed a simple methodology for Type Ia supernova cosmology that accounts for Malmquist bias using simulation-based inference and hierarchical Bayesian modelling. This method is unique compared to other simulation-based inference applications in the field, because it can infer different sets of cosmological parameters and is robust to varying SN sample sizes, without needing to be retrained. The likelihood to be learnt is only over three dimensions, does not require data compression beyond the three SN light curve summary parameters and importantly is trained independent of cosmology. These factors allow the likelihood to be easier to learn in a domain where simulations may be expensive. 

We validate our method on a toy simulation that has an analytical likelihood solution. Results showed that the method was able to recover true simulated cosmology and most supernova hyperparameters and agreed with the analytical posteriors to within $1 \sigma$. The method provided almost identical posteriors on supernova hyperparameters when inferring parameters from two different cosmological models. 

This technique can be extended to provide useful insights on the astrophysics of SNe as it is flexible enough to incorporate intrinsic and extrinsic host galaxy environment effects. Furthermore, we are training the model on state-of-the-art survey simulations to learn non-analytical likelihoods caused by more complex selection effects. These learnt likelihoods may be used in future hierarchical cosmological analyses combining multiple surveys to further constrain the nature of our universe. 

\newpage

%\section*{Software and Data}
%If a paper is accepted, we strongly encourage the publication of software and data with the
%camera-ready version of the paper whenever appropriate. This can be
%done by including a URL in the camera-ready copy. However, \textbf{do not}
%include URLs that reveal your institution or identity in your
%submission for review. Instead, provide an anonymous URL or upload
%the material as ``Supplementary Material'' into the CMT reviewing
%system. Note that reviewers are not required to look at this material
%when writing their review.

% Acknowledgements should only appear in the accepted version.
\section*{Acknowledgements}
BMB is supported by the Cambridge Centre for Doctoral Training in Data-Intensive Science funded by the UK Science and Technology Facilities Council (STFC) and two G-Research early career researchers grants used for equipment and travel. MG and KSM are supported by the European Union’s Horizon 2020
research and innovation programme under European Research Council (ERC) Grant Agreement No. 101002652 and Marie Skłodowska Curie Grant Agreement No. 873089. ST was supported by the European Research Council (ERC) under the European Union's Horizon 2020 research and innovation programme (grant agreement no.\ 101018897 CosmicExplorer).
%\textbf{Do not} include acknowledgements in the initial version of
%the paper submitted for blind review.

%If a paper is accepted, the final camera-ready version can (and
%probably should) include acknowledgements. In this case, please
%place such acknowledgements in an unnumbered section at the
%end of the paper. Typically, this will include thanks to reviewers
%who gave useful comments, to colleagues who contributed to the ideas,
%and to funding agencies and corporate sponsors that provided financial
%support.

% In the unusual situation where you want a paper to appear in the
% references without citing it in the main text, use \nocite

\bibliography{main_}

\begin{thebibliography}{36}
\providecommand{\natexlab}[1]{#1}
\providecommand{\url}[1]{\texttt{#1}}
\expandafter\ifx\csname urlstyle\endcsname\relax
  \providecommand{\doi}[1]{doi: #1}\else
  \providecommand{\doi}{doi: \begingroup \urlstyle{rm}\Url}\fi

\bibitem[{Alsing} et~al.(2019){Alsing}, {Charnock}, {Feeney}, and {Wandelt}]{alsing2019}
{Alsing}, J., {Charnock}, T., {Feeney}, S., and {Wandelt}, B.
\newblock {Fast likelihood-free cosmology with neural density estimators and active learning}.
\newblock \emph{\mnras}, 488\penalty0 (3):\penalty0 4440--4458, September 2019.
\newblock \doi{10.1093/mnras/stz1960}.

\bibitem[Bingham et~al.(2019)Bingham, Chen, Jankowiak, Obermeyer, Pradhan, Karaletsos, Singh, Szerlip, Horsfall, and Goodman]{pyro}
Bingham, E., Chen, J.~P., Jankowiak, M., Obermeyer, F., Pradhan, N., Karaletsos, T., Singh, R., Szerlip, P.~A., Horsfall, P., and Goodman, N.~D.
\newblock Pyro: Deep universal probabilistic programming.
\newblock \emph{J. Machine Learning Res.}, 20:\penalty0 28:1--28:6, 2019.
\newblock URL \url{http://jmlr.org/papers/v20/18-403.html}.

\bibitem[Bishop(1994)]{bishop1994}
Bishop, C.~M.
\newblock Mixture density networks.
\newblock Technical Report NCRG/94/004, Aston University, 1994.
\newblock URL \url{https://publications.aston.ac.uk/id/eprint/373/1/NCRG_94_004.pdf}.

\bibitem[Bradbury et~al.(2018)Bradbury, Frostig, Hawkins, Johnson, Leary, Maclaurin, Necula, Paszke, Vander{P}las, Wanderman-{M}ilne, and Zhang]{jax2018github}
Bradbury, J., Frostig, R., Hawkins, P., Johnson, M.~J., Leary, C., Maclaurin, D., Necula, G., Paszke, A., Vander{P}las, J., Wanderman-{M}ilne, S., and Zhang, Q.
\newblock {JAX}: composable transformations of {P}ython+{N}um{P}y programs, 2018.
\newblock URL \url{http://github.com/google/jax}.

\bibitem[{Campagne} et~al.(2023){Campagne}, {Lanusse}, {Zuntz}, {Boucaud}, {Casas}, {Karamanis}, {Kirkby}, {Lanzieri}, {Peel}, and {Li}]{jaxcosmo}
{Campagne}, J.-E., {Lanusse}, F., {Zuntz}, J., {Boucaud}, A., {Casas}, S., {Karamanis}, M., {Kirkby}, D., {Lanzieri}, D., {Peel}, A., and {Li}, Y.
\newblock {JAX-COSMO: An End-to-End Differentiable and GPU Accelerated Cosmology Library}.
\newblock \emph{Open J.\ Astrophys.}, 6:\penalty0 15, April 2023.
\newblock \doi{10.21105/astro.2302.05163}.

\bibitem[{Carr} et~al.(2022){Carr}, {Davis}, {Scolnic}, {Said}, {Brout}, {Peterson}, and {Kessler}]{carr2022}
{Carr}, A., {Davis}, T.~M., {Scolnic}, D., {Said}, K., {Brout}, D., {Peterson}, E.~R., and {Kessler}, R.
\newblock {The Pantheon+ analysis: Improving the redshifts and peculiar velocities of Type Ia supernovae used in cosmological analyses}.
\newblock \emph{\pasa}, 39:\penalty0 e046, October 2022.
\newblock \doi{10.1017/pasa.2022.41}.

\bibitem[{Childress} et~al.(2013){Childress}, {Aldering}, {Antilogus}, {Aragon}, {Bailey}, {Baltay}, {Bongard}, {Buton}, {Canto}, {Cellier-Holzem}, {Chotard}, {Copin}, {Fakhouri}, {Gangler}, {Guy}, {Hsiao}, {Kerschhaggl}, {Kim}, {Kowalski}, {Loken}, {Nugent}, {Paech}, {Pain}, {Pecontal}, {Pereira}, {Perlmutter}, {Rabinowitz}, {Rigault}, {Runge}, {Scalzo}, {Smadja}, {Tao}, {Thomas}, {Weaver}, and {Wu}]{childress2013}
{Childress}, M., {Aldering}, G., {Antilogus}, P., {Aragon}, C., {Bailey}, S., {Baltay}, C., {Bongard}, S., {Buton}, C., {Canto}, A., {Cellier-Holzem}, F., {Chotard}, N., {Copin}, Y., {Fakhouri}, H.~K., {Gangler}, E., {Guy}, J., {Hsiao}, E.~Y., {Kerschhaggl}, M., {Kim}, A.~G., {Kowalski}, M., {Loken}, S., {Nugent}, P., {Paech}, K., {Pain}, R., {Pecontal}, E., {Pereira}, R., {Perlmutter}, S., {Rabinowitz}, D., {Rigault}, M., {Runge}, K., {Scalzo}, R., {Smadja}, G., {Tao}, C., {Thomas}, R.~C., {Weaver}, B.~A., and {Wu}, C.
\newblock {Host Galaxy Properties and Hubble Residuals of Type Ia Supernovae from the Nearby Supernova Factory}.
\newblock \emph{\apj}, 770\penalty0 (2):\penalty0 108, June 2013.
\newblock \doi{10.1088/0004-637X/770/2/108}.

\bibitem[Duane et~al.(1987)Duane, Kennedy, Pendleton, and Roweth]{duane1987}
Duane, S., Kennedy, A., Pendleton, B.~J., and Roweth, D.
\newblock Hybrid monte carlo.
\newblock \emph{Phys.\ Lett.\ B}, 195\penalty0 (2):\penalty0 216--222, 1987.
\newblock ISSN 0370-2693.
\newblock \doi{10.1016/0370-2693(87)91197-X}.

\bibitem[Germain et~al.(2015)Germain, Gregor, Murray, and Larochelle]{germain2015made}
Germain, M., Gregor, K., Murray, I., and Larochelle, H.
\newblock Made: Masked autoencoder for distribution estimation.
\newblock In Bach, F. and Blei, D. (eds.), \emph{Proceedings of the 32nd International Conference on Machine Learning}, volume~37 of \emph{Proceedings of Machine Learning Research}, pp.\  881--889, Lille, France, 07--09 Jul 2015. PMLR.
\newblock URL \url{https://proceedings.mlr.press/v37/germain15.html}.

\bibitem[{Guy} et~al.(2007){Guy}, {Astier}, {Baumont}, {Hardin}, {Pain}, {Regnault}, {Basa}, {Carlberg}, {Conley}, {Fabbro}, {Fouchez}, {Hook}, {Howell}, {Perrett}, {Pritchet}, {Rich}, {Sullivan}, {Antilogus}, {Aubourg}, {Bazin}, {Bronder}, {Filiol}, {Palanque-Delabrouille}, {Ripoche}, and {Ruhlmann-Kleider}]{guy2007}
{Guy}, J., {Astier}, P., {Baumont}, S., {Hardin}, D., {Pain}, R., {Regnault}, N., {Basa}, S., {Carlberg}, R.~G., {Conley}, A., {Fabbro}, S., {Fouchez}, D., {Hook}, I.~M., {Howell}, D.~A., {Perrett}, K., {Pritchet}, C.~J., {Rich}, J., {Sullivan}, M., {Antilogus}, P., {Aubourg}, E., {Bazin}, G., {Bronder}, J., {Filiol}, M., {Palanque-Delabrouille}, N., {Ripoche}, P., and {Ruhlmann-Kleider}, V.
\newblock {SALT2: using distant supernovae to improve the use of type Ia supernovae as distance indicators}.
\newblock \emph{\aap}, 466\penalty0 (1):\penalty0 11--21, April 2007.
\newblock \doi{10.1051/0004-6361:20066930}.

\bibitem[{Guy} et~al.(2010){Guy}, {Sullivan}, {Conley}, {Regnault}, {Astier}, {Balland}, {Basa}, {Carlberg}, {Fouchez}, {Hardin}, {Hook}, {Howell}, {Pain}, {Palanque-Delabrouille}, {Perrett}, {Pritchet}, {Rich}, {Ruhlmann-Kleider}, {Balam}, {Baumont}, {Ellis}, {Fabbro}, {Fakhouri}, {Fourmanoit}, {Gonz{\'a}lez-Gait{\'a}n}, {Graham}, {Hsiao}, {Kronborg}, {Lidman}, {Mourao}, {Perlmutter}, {Ripoche}, {Suzuki}, and {Walker}]{guy2010}
{Guy}, J., {Sullivan}, M., {Conley}, A., {Regnault}, N., {Astier}, P., {Balland}, C., {Basa}, S., {Carlberg}, R.~G., {Fouchez}, D., {Hardin}, D., {Hook}, I.~M., {Howell}, D.~A., {Pain}, R., {Palanque-Delabrouille}, N., {Perrett}, K.~M., {Pritchet}, C.~J., {Rich}, J., {Ruhlmann-Kleider}, V., {Balam}, D., {Baumont}, S., {Ellis}, R.~S., {Fabbro}, S., {Fakhouri}, H.~K., {Fourmanoit}, N., {Gonz{\'a}lez-Gait{\'a}n}, S., {Graham}, M.~L., {Hsiao}, E., {Kronborg}, T., {Lidman}, C., {Mourao}, A.~M., {Perlmutter}, S., {Ripoche}, P., {Suzuki}, N., and {Walker}, E.~S.
\newblock {The Supernova Legacy Survey 3-year sample: Type Ia supernovae photometric distances and cosmological constraints}.
\newblock \emph{\aap}, 523:\penalty0 A7, November 2010.
\newblock \doi{10.1051/0004-6361/201014468}.

\bibitem[Hermans et~al.(2020)Hermans, Begy, and Louppe]{hermans2020}
Hermans, J., Begy, V., and Louppe, G.
\newblock Likelihood-free {MCMC} with amortized approximate ratio estimators.
\newblock In Daum\'{e}~III, H. and Singh, A. (eds.), \emph{Proceedings of the 37th International Conference on Machine Learning}, volume 119 of \emph{Proceedings of Machine Learning Research}, pp.\  4239--4248. PMLR, 13--18 Jul 2020.
\newblock URL \url{https://proceedings.mlr.press/v119/hermans20a.html}.

\bibitem[{Karchev} et~al.(2023{\natexlab{a}}){Karchev}, {Trotta}, and {Weniger}]{karchev2023}
{Karchev}, K., {Trotta}, R., and {Weniger}, C.
\newblock {SICRET: Supernova Ia Cosmology with truncated marginal neural Ratio EsTimation}.
\newblock \emph{\mnras}, 520\penalty0 (1):\penalty0 1056--1072, March 2023{\natexlab{a}}.
\newblock \doi{10.1093/mnras/stac3785}.

\bibitem[{Karchev} et~al.(2023{\natexlab{b}}){Karchev}, {Trotta}, and {Weniger}]{karchev2023b}
{Karchev}, K., {Trotta}, R., and {Weniger}, C.
\newblock {SimSIMS: Simulation-based Supernova Ia Model Selection with thousands of latent variables}.
\newblock \emph{arXiv e-prints}, art. arXiv:2311.15650, November 2023{\natexlab{b}}.
\newblock \doi{10.48550/arXiv.2311.15650}.

\bibitem[{Karchev} et~al.(2024){Karchev}, {Grayling}, {Boyd}, {Trotta}, {Mandel}, and {Weniger}]{karchev2024}
{Karchev}, K., {Grayling}, M., {Boyd}, B.~M., {Trotta}, R., {Mandel}, K.~S., and {Weniger}, C.
\newblock {SIDE-real: Supernova Ia Dust Extinction with truncated marginal neural ratio estimation applied to real data}.
\newblock \emph{\mnras}, 530\penalty0 (4):\penalty0 3881--3896, June 2024.
\newblock \doi{10.1093/mnras/stae995}.

\bibitem[{Kelly} et~al.(2010){Kelly}, {Hicken}, {Burke}, {Mandel}, and {Kirshner}]{kelly2010}
{Kelly}, P.~L., {Hicken}, M., {Burke}, D.~L., {Mandel}, K.~S., and {Kirshner}, R.~P.
\newblock {Hubble Residuals of Nearby Type Ia Supernovae are Correlated with Host Galaxy Masses}.
\newblock \emph{\apj}, 715\penalty0 (2):\penalty0 743--756, June 2010.
\newblock \doi{10.1088/0004-637X/715/2/743}.

\bibitem[{Kessler} \& {Scolnic}(2017){Kessler} and {Scolnic}]{kessler2017}
{Kessler}, R. and {Scolnic}, D.
\newblock {Correcting Type Ia Supernova Distances for Selection Biases and Contamination in Photometrically Identified Samples}.
\newblock \emph{\apj}, 836\penalty0 (1):\penalty0 56, February 2017.
\newblock \doi{10.3847/1538-4357/836/1/56}.

\bibitem[{Kessler} et~al.(2009){Kessler}, {Bernstein}, {Cinabro}, {Dilday}, {Frieman}, {Jha}, {Kuhlmann}, {Miknaitis}, {Sako}, {Taylor}, and {Vanderplas}]{kessler2009}
{Kessler}, R., {Bernstein}, J.~P., {Cinabro}, D., {Dilday}, B., {Frieman}, J.~A., {Jha}, S., {Kuhlmann}, S., {Miknaitis}, G., {Sako}, M., {Taylor}, M., and {Vanderplas}, J.
\newblock {SNANA: A Public Software Package for Supernova Analysis}.
\newblock \emph{\pasp}, 121\penalty0 (883):\penalty0 1028, September 2009.
\newblock \doi{10.1086/605984}.

\bibitem[{Lampeitl} et~al.(2010){Lampeitl}, {Smith}, {Nichol}, {Bassett}, {Cinabro}, {Dilday}, {Foley}, {Frieman}, {Garnavich}, {Goobar}, {Im}, {Jha}, {Marriner}, {Miquel}, {Nordin}, {{\"O}stman}, {Riess}, {Sako}, {Schneider}, {Sollerman}, and {Stritzinger}]{lampeit2010}
{Lampeitl}, H., {Smith}, M., {Nichol}, R.~C., {Bassett}, B., {Cinabro}, D., {Dilday}, B., {Foley}, R.~J., {Frieman}, J.~A., {Garnavich}, P.~M., {Goobar}, A., {Im}, M., {Jha}, S.~W., {Marriner}, J., {Miquel}, R., {Nordin}, J., {{\"O}stman}, L., {Riess}, A.~G., {Sako}, M., {Schneider}, D.~P., {Sollerman}, J., and {Stritzinger}, M.
\newblock {The Effect of Host Galaxies on Type Ia Supernovae in the SDSS-II Supernova Survey}.
\newblock \emph{\apj}, 722\penalty0 (1):\penalty0 566--576, October 2010.
\newblock \doi{10.1088/0004-637X/722/1/566}.

\bibitem[{Malmquist}(1922)]{malmquist1922}
{Malmquist}, K.~G.
\newblock {On some relations in stellar statistics}.
\newblock \emph{Medd.\ Lunds Astron.\ Obser.\ Serie I}, 100:\penalty0 1--52, March 1922.

\bibitem[{Mandel} et~al.(2009){Mandel}, {Wood-Vasey}, {Friedman}, and {Kirshner}]{mandel2009}
{Mandel}, K.~S., {Wood-Vasey}, W.~M., {Friedman}, A.~S., and {Kirshner}, R.~P.
\newblock {Type Ia Supernova Light-Curve Inference: Hierarchical Bayesian Analysis in the Near-Infrared}.
\newblock \emph{\apj}, 704\penalty0 (1):\penalty0 629--651, October 2009.
\newblock \doi{10.1088/0004-637X/704/1/629}.

\bibitem[{Mandel} et~al.(2017){Mandel}, {Scolnic}, {Shariff}, {Foley}, and {Kirshner}]{mandel2017}
{Mandel}, K.~S., {Scolnic}, D.~M., {Shariff}, H., {Foley}, R.~J., and {Kirshner}, R.~P.
\newblock {The Type Ia Supernova Color-Magnitude Relation and Host Galaxy Dust: A Simple Hierarchical Bayesian Model}.
\newblock \emph{\apj}, 842\penalty0 (2):\penalty0 93, June 2017.
\newblock \doi{10.3847/1538-4357/aa6038}.

\bibitem[{March} et~al.(2011){March}, {Trotta}, {Berkes}, {Starkman}, and {Vaudrevange}]{march2011}
{March}, M.~C., {Trotta}, R., {Berkes}, P., {Starkman}, G.~D., and {Vaudrevange}, P.~M.
\newblock {Improved constraints on cosmological parameters from Type Ia supernova data}.
\newblock \emph{\mnras}, 418\penalty0 (4):\penalty0 2308--2329, December 2011.
\newblock \doi{10.1111/j.1365-2966.2011.19584.x}.

\bibitem[{March} et~al.(2018){March}, {Wolf}, {Sako}, {D'Andrea}, and {Brout}]{march2018}
{March}, M.~C., {Wolf}, R.~C., {Sako}, m., {D'Andrea}, C., and {Brout}, D.
\newblock {A Bayesian approach to truncated data sets: An application to Malmquist bias in Supernova Cosmology}.
\newblock \emph{arXiv e-prints}, art. arXiv:1804.02474, April 2018.
\newblock \doi{10.48550/arXiv.1804.02474}.

\bibitem[Papamakarios et~al.(2017)Papamakarios, Pavlakou, and Murray]{papamakarios2017masked}
Papamakarios, G., Pavlakou, T., and Murray, I.
\newblock Masked autoregressive flow for density estimation.
\newblock In Guyon, I., Luxburg, U.~V., Bengio, S., Wallach, H., Fergus, R., Vishwanathan, S., and Garnett, R. (eds.), \emph{Advances in Neural Information Processing Systems}, volume~30. Curran Associates, Inc., 2017.
\newblock URL \url{https://proceedings.neurips.cc/paper_files/paper/2017/file/6c1da886822c67822bcf3679d04369fa-Paper.pdf}.

\bibitem[Papamakarios et~al.(2021)Papamakarios, Nalisnick, Rezende, Mohamed, and Lakshminarayanan]{papamakarios2021}
Papamakarios, G., Nalisnick, E., Rezende, D.~J., Mohamed, S., and Lakshminarayanan, B.
\newblock Normalizing flows for probabilistic modeling and inference.
\newblock \emph{J.\ Machine Learning Res.}, 22\penalty0 (57):\penalty0 1--64, 2021.
\newblock URL \url{http://jmlr.org/papers/v22/19-1028.html}.

\bibitem[{Phan} et~al.(2019){Phan}, {Pradhan}, and {Jankowiak}]{numpyro}
{Phan}, D., {Pradhan}, N., and {Jankowiak}, M.
\newblock {Composable Effects for Flexible and Accelerated Probabilistic Programming in NumPyro}.
\newblock \emph{arXiv e-prints}, art. arXiv:1912.11554, December 2019.
\newblock \doi{10.48550/arXiv.1912.11554}.

\bibitem[Rezende \& Mohamed(2015)Rezende and Mohamed]{rezende2015}
Rezende, D. and Mohamed, S.
\newblock Variational inference with normalizing flows.
\newblock In Bach, F. and Blei, D. (eds.), \emph{Proceedings of the 32nd International Conference on Machine Learning}, volume~37 of \emph{Proceedings of Machine Learning Research}, pp.\  1530--1538, Lille, France, 07--09 Jul 2015. PMLR.
\newblock URL \url{https://proceedings.mlr.press/v37/rezende15.html}.

\bibitem[{Rubin} et~al.(2015){Rubin}, {Aldering}, {Barbary}, {Boone}, {Chappell}, {Currie}, {Deustua}, {Fagrelius}, {Fruchter}, {Hayden}, {Lidman}, {Nordin}, {Perlmutter}, {Saunders}, {Sofiatti}, and {Supernova Cosmology Project}]{rubin2015}
{Rubin}, D., {Aldering}, G., {Barbary}, K., {Boone}, K., {Chappell}, G., {Currie}, M., {Deustua}, S., {Fagrelius}, P., {Fruchter}, A., {Hayden}, B., {Lidman}, C., {Nordin}, J., {Perlmutter}, S., {Saunders}, C., {Sofiatti}, C., and {Supernova Cosmology Project}, T.
\newblock {UNITY: Confronting Supernova Cosmology's Statistical and Systematic Uncertainties in a Unified Bayesian Framework}.
\newblock \emph{\apj}, 813\penalty0 (2):\penalty0 137, November 2015.
\newblock \doi{10.1088/0004-637X/813/2/137}.

\bibitem[{Rubin} et~al.(2023){Rubin}, {Aldering}, {Betoule}, {Fruchter}, {Huang}, {Kim}, {Lidman}, {Linder}, {Perlmutter}, {Ruiz-Lapuente}, and {Suzuki}]{rubin2023}
{Rubin}, D., {Aldering}, G., {Betoule}, M., {Fruchter}, A., {Huang}, X., {Kim}, A.~G., {Lidman}, C., {Linder}, E., {Perlmutter}, S., {Ruiz-Lapuente}, P., and {Suzuki}, N.
\newblock {Union Through UNITY: Cosmology with 2,000 SNe Using a Unified Bayesian Framework}.
\newblock \emph{arXiv e-prints}, art. arXiv:2311.12098, November 2023.
\newblock \doi{10.48550/arXiv.2311.12098}.

\bibitem[{Scolnic} et~al.(2022){Scolnic}, {Brout}, {Carr}, {Riess}, {Davis}, {Dwomoh}, {Jones}, {Ali}, {Charvu}, {Chen}, {Peterson}, {Popovic}, {Rose}, {Wood}, {Brown}, {Chambers}, {Coulter}, {Dettman}, {Dimitriadis}, {Filippenko}, {Foley}, {Jha}, {Kilpatrick}, {Kirshner}, {Pan}, {Rest}, {Rojas-Bravo}, {Siebert}, {Stahl}, and {Zheng}]{scolnic2022}
{Scolnic}, D., {Brout}, D., {Carr}, A., {Riess}, A.~G., {Davis}, T.~M., {Dwomoh}, A., {Jones}, D.~O., {Ali}, N., {Charvu}, P., {Chen}, R., {Peterson}, E.~R., {Popovic}, B., {Rose}, B.~M., {Wood}, C.~M., {Brown}, P.~J., {Chambers}, K., {Coulter}, D.~A., {Dettman}, K.~G., {Dimitriadis}, G., {Filippenko}, A.~V., {Foley}, R.~J., {Jha}, S.~W., {Kilpatrick}, C.~D., {Kirshner}, R.~P., {Pan}, Y.-C., {Rest}, A., {Rojas-Bravo}, C., {Siebert}, M.~R., {Stahl}, B.~E., and {Zheng}, W.
\newblock {The Pantheon+ Analysis: The Full Data Set and Light-curve Release}.
\newblock \emph{\apj}, 938\penalty0 (2):\penalty0 113, October 2022.
\newblock \doi{10.3847/1538-4357/ac8b7a}.

\bibitem[{Shariff} et~al.(2016){Shariff}, {Jiao}, {Trotta}, and {van Dyk}]{shariff2016}
{Shariff}, H., {Jiao}, X., {Trotta}, R., and {van Dyk}, D.~A.
\newblock {BAHAMAS: New Analysis of Type Ia Supernovae Reveals Inconsistencies with Standard Cosmology}.
\newblock \emph{\apj}, 827\penalty0 (1):\penalty0 1, August 2016.
\newblock \doi{10.3847/0004-637X/827/1/1}.

\bibitem[{Sullivan} et~al.(2010){Sullivan}, {Conley}, {Howell}, {Neill}, {Astier}, {Balland}, {Basa}, {Carlberg}, {Fouchez}, {Guy}, {Hardin}, {Hook}, {Pain}, {Palanque-Delabrouille}, {Perrett}, {Pritchet}, {Regnault}, {Rich}, {Ruhlmann-Kleider}, {Baumont}, {Hsiao}, {Kronborg}, {Lidman}, {Perlmutter}, and {Walker}]{sullivan2010}
{Sullivan}, M., {Conley}, A., {Howell}, D.~A., {Neill}, J.~D., {Astier}, P., {Balland}, C., {Basa}, S., {Carlberg}, R.~G., {Fouchez}, D., {Guy}, J., {Hardin}, D., {Hook}, I.~M., {Pain}, R., {Palanque-Delabrouille}, N., {Perrett}, K.~M., {Pritchet}, C.~J., {Regnault}, N., {Rich}, J., {Ruhlmann-Kleider}, V., {Baumont}, S., {Hsiao}, E., {Kronborg}, T., {Lidman}, C., {Perlmutter}, S., and {Walker}, E.~S.
\newblock {The dependence of Type Ia Supernovae luminosities on their host galaxies}.
\newblock \emph{\mnras}, 406\penalty0 (2):\penalty0 782--802, August 2010.
\newblock \doi{10.1111/j.1365-2966.2010.16731.x}.

\bibitem[Tabak \& Turner(2013)Tabak and Turner]{tabak2013family}
Tabak, E.~G. and Turner, C.~V.
\newblock A family of nonparametric density estimation algorithms.
\newblock \emph{Comm. Pure Applied Math.}, 66\penalty0 (2):\penalty0 145--164, 2013.

\bibitem[Tabak \& Vanden-Eijnden(2010)Tabak and Vanden-Eijnden]{tabak2010density}
Tabak, E.~G. and Vanden-Eijnden, E.
\newblock Density estimation by dual ascent of the log-likelihood.
\newblock \emph{Comm. Math. Sci.}, 8\penalty0 (1):\penalty0 217--233, 2010.

\bibitem[{Tripp}(1998)]{tripp1998}
{Tripp}, R.
\newblock {A two-parameter luminosity correction for Type IA supernovae}.
\newblock \emph{\aap}, 331:\penalty0 815--820, March 1998.

\end{thebibliography}

\bibliographystyle{icml2021}

\newpage

\onecolumn
\icmltitle{Supplementary Material:\\
Accounting for Selection Effects in Supernova Cosmology with Simulation-Based Inference and Hierarchical Bayesian Modelling}
\appendix
\section{Analytical Solution to Toy Model}
\label{sec:proof}
To analytically write the likelihood for the toy model outlined in Section \ref{sec:toy}, we need to solve the integral in the denominator of Equation \eqref{eq:genliknew}. To begin we use the conditional properties of the multivariate Gaussian to write: 
\begin{equation}
N\big(\bm{\hat{d}_{s}}\big|\,\bm{\Psi_{d}},\bm{\Sigma_d}\big) = N\big(\hat{m}_s \big|\, \text{E}[\hat{m}_s|\,\hat{c}_s,\hat{x}_s],\text{Var} [\hat{m}_s|\,\hat{c}_s,\hat{x}_s]\big)N\big(\hat{c}_s \big|\, \text{E}[\hat{c}_s|\,\hat{x}_s],\text{Var} [\hat{c}_s|\,\hat{x}_s]\big)N\big(\hat{x}_s \big|\, \text{E}[\hat{x}_s],\text{Var} [\hat{x}_s]\big)
\end{equation}
where 
\begin{align*}
    \text{E}[\hat{c}_s|\,\hat{x}_s]&= \text{E}[\hat{c}_s] + \frac{\text{Cov}[\hat{c}_s,\hat{x}_s]}{\text{Var}[\hat{x}_s]}\big(\hat{x}_s -  \text{E}[\hat{x}_s]\big) \\
 \text{Var}[\hat{c}_s|\,\hat{x}_s]&= \text{Var}[\hat{c}_s] - \frac{\text{Cov}[\hat{c},\hat{x}]^2}{\text{Var}[\hat{x}_s]} \\   
\text{E}[\hat{m}_s|\,\hat{x}_s] &= \text{E}[\hat{m}_s] + \frac{\text{Cov}[\hat{m}_s,\hat{x}_s]}{\text{Var}[\hat{x}_s]}\big(\hat{x}_s -  \text{E}[\hat{x}_s]\big)
 \\
 \text{Var}[\hat{m}_s|\,\hat{x}_s] &= \text{Var}[\hat{m}_s] - \frac{\text{Cov}[\hat{m},\hat{x}]^2}{\text{Var}[\hat{x}_s]}\\
\text{Cov}[\hat{m}_s,\hat{c}_s|\,\hat{x}_s]&= \text{Cov}[\hat{m}_s,\hat{c}_s]-\frac{\text{Cov}[\hat{m}_s,\hat{x}_s]\text{Cov}[\hat{c}_s,\hat{x}_s]}{\text{Var}[\hat{x}]}
 \\
     \text{E}[\hat{m}_s|\,\hat{c}_s,\hat{x}_s] &= \text{E}[\hat{m}_s|\,\hat{x}_s] + \frac{\text{Cov}[\hat{m}_s,\hat{c}_s|\,\hat{x}_s]}{ \text{Var}[\hat{c}_s|\,\hat{x}_s]}\big(\hat{c}_s -  \text{E}[\hat{c}_s|\,\hat{x}_s]\big)\\
\text{Var}[\hat{m}_s|\,\hat{c}_s,\hat{x}_s] &= \text{Var}[\hat{m}_s|\,\hat{x}_s] - \frac{\text{Cov}[\hat{m}_s,\hat{c}_s|\,\hat{x}_s]^2}{ \text{Var}[\hat{c}_s|\,\hat{x}_s]}
\end{align*}
Marginal expectations and covariances are defined in Section \ref{sec:toy}.

We can then solve the integral using Equation \eqref{eq:indentity} by integrating each dimension one at a time with the selection function:
\begin{multline}
   \int^{\infty}_{-\infty}\Phi \Bigg(\frac{m_{\text{cut}} -(\hat{m}_s + a_{\text{cut}} \hat{x}_s + b_{\text{cut}} \hat{c}_s)}{\sigma_{\text{cut}}}\Bigg)  N\big(\bm{\hat{d}_{s}}\big|\,\bm{\Psi_{d}},\bm{\Sigma_d}\big) \,\text{d}\hat{d}_s \\
    \hspace{-6cm}= \int^{\infty}_{-\infty} \int^{\infty}_{-\infty} \int^{\infty}_{-\infty} \Bigg[\Phi \Bigg(\frac{m_{\text{cut}} -(\hat{m}_s + a_{\text{cut}} \hat{x}_s + b_{\text{cut}} \hat{c}_s)}{\sigma_{\text{cut}}}\Bigg) \times \\
   N\big(\hat{m}_s \big|\, \text{E}[\hat{m}_s|\,\hat{c}_s,\hat{x}_s],\text{Var} [\hat{m}_s|\,\hat{c}_s,\hat{x}_s]\big)N\big(\hat{c}_s \big|\, \text{E}[\hat{c}_s|\,\hat{x}_s],\text{Var} [\hat{c}_s|\,\hat{x}_s]\big)N\big(\hat{x}_s \big|\, \text{E}[\hat{x}_s],\text{Var} [\hat{x}_s]\big)\Bigg]\,\text{d}\hat{m}_s\,\text{d}\hat{c}_s\,\text{d}\hat{x}_s \\
\end{multline}
The first dimension $\hat{m}_s$ is integrated to get: 
\begin{multline}
    \int^{\infty}_{-\infty} \Phi \Bigg(\frac{m_{\text{cut}} -(\hat{m}_s + a_{\text{cut}} \hat{x}_s + b_{\text{cut}} \hat{c}_s)}{\sigma_{\text{cut}}}\Bigg) 
   N\big(\hat{m}_s \big|\, \text{E}[\hat{m}_s|\,\hat{c}_s,\hat{x}_s],\text{Var} [\hat{m}_s|\,\hat{c}_s,\hat{x}_s]\big)\,\text{d}\hat{m}_s \\
= \Phi \Bigg(\frac{m_{\text{cut}} -( \text{E}[\hat{m}_s|\,\hat{c}_s,\hat{x}_s] + a_{\text{cut}} \hat{x}_s + b_{\text{cut}} \hat{c}_s)}{\sigma'_{\text{cut}}}\Bigg) 
\end{multline}
where 
\begin{equation*}
    \sigma'_{cut}= \sigma_{cut}\sqrt{1+ \frac{\text{Var}[\hat{m}_s|\,\hat{c}_s,\hat{x}_s]}{\sigma_{cut}^2}}
\end{equation*}
The second dimension $\hat{c}_s$
is then integrated out to get:
\begin{multline}
    \int^{\infty}_{-\infty} \Phi \Bigg(\frac{m_{\text{cut}} -(\text{E}[\hat{m}_s|\,\hat{c}_s,\hat{x}_s] +a_{\text{cut}} \hat{x}_s + b_{\text{cut}} \hat{c}_s)}{\sigma'_{\text{cut}}}\Bigg) 
   N\big(\hat{c}_s \big|\, \text{E}[\hat{c}_s|\,\hat{x}_s],\text{Var} [\hat{c}_s|\,\hat{x}_s]\big)\,\text{d}\hat{c}_s \\   = \int^{\infty}_{-\infty} \Phi \Bigg(\frac{m_{\text{cut}} -(\text{E}[\hat{m}_s|\,\hat{x}_s]+\gamma_c (\hat{c}_s - \text{E}[\hat{c}_s|\,\hat{x}_s]) +a_{\text{cut}} \hat{x}_s + b_{\text{cut}} \hat{c}_s)}{\sigma'_{\text{cut}}}\Bigg) 
   N\big(\hat{c}_s \big|\, \text{E}[\hat{c}_s|\,\hat{x}_s],\text{Var} [\hat{c}_s|\,\hat{x}_s]\big)\,\text{d}\hat{c}_s \\
= \Phi \Bigg(\frac{m_{\text{cut}} -( \text{E}[\hat{m}_s|\,\hat{x}_s] -\gamma_c \text{E}[\hat{c}_s|\,\hat{x}_s] +a_{\text{cut}} \hat{x}_s + (b_{\text{cut}} + \gamma_c )\text{E}[\hat{c}_s|\,\hat{x}_s])}{\sigma''_{\text{cut}}}\Bigg) \\
= \Phi \Bigg(\frac{m_{\text{cut}} -( \text{E}[\hat{m}_s|\,\hat{x}_s] +a_{\text{cut}} \hat{x}_s + b_{\text{cut}}\text{E}[\hat{c}_s|\,\hat{x}_s])}{\sigma''_{\text{cut}}}\Bigg) 
\end{multline}
where 
\begin{align*}
    \sigma''_{cut}&= \sigma'_{cut}\sqrt{1+ \frac{(b_{\text{cut}} + \gamma_c)^2\text{Var}[\hat{c}_s|\,\hat{x}_s]}{{\sigma'_{cut}}^2}}
\\
\gamma_c &= \frac{\text{Cov}[\hat{m}_s,\hat{c}_s|\,\hat{x}_s]}{\text{Var}[\hat{c}_s|\,\hat{x}_s]}
\end{align*}
Then the third dimension $\hat{x}_s$ is integrated to get the final result:
\begin{multline}
    \int^{\infty}_{-\infty} \Phi \Bigg(\frac{m_{\text{cut}} -( \text{E}[\hat{m}_s|\,\hat{x}_s] +a_{\text{cut}} \hat{x}_s + b_{\text{cut}}\text{E}[\hat{c}_s|\,\hat{x}_s])}{\sigma''_{\text{cut}}}\Bigg)  
   N\big(\hat{x}_s \big|\, \text{E}[\hat{x}_s],\text{Var} [\hat{x}_s]\big)\,\text{d}\hat{x}_s\\=
       \int^{\infty}_{-\infty} \Phi \Bigg(\frac{m_{\text{cut}} -( \text{E}[\hat{m}_s]+\gamma_x(\hat{x}_s -\text{E}[\hat{x}_s]) +a_{\text{cut}} \hat{x}_s + b_{\text{cut}}(\text{E}[\hat{c}_s]+\delta_x(\hat{x}_s -\text{E}[\hat{x}_s]))}{\sigma''_{\text{cut}}}\Bigg)  
   N\big(\hat{x}_s \big|\, \text{E}[\hat{x}_s],\text{Var} [\hat{x}_s]\big)\,\text{d}\hat{x}_s\\
= \Phi \Bigg(\frac{m_{\text{cut}} -( \text{E}[\hat{m}_s] - \gamma_x \text{E}[\hat{x}] +(a_{\text{cut}}+\gamma_x + \delta_x b_{\text{cut}} )\text{E}[\hat{x}_s]+ b_{\text{cut}} (\text{E}[\hat{c}_s]-\delta_x E[\hat{x}])}{\sigma'''_{\text{cut}}}\Bigg) \\
=\Phi \Bigg(\frac{m_{\text{cut}} -( \text{E}[\hat{m}_s] +a_{\text{cut}} \text{E}[\hat{x}_s]+ b_{\text{cut}} \text{E}[\hat{c}_s])}{\sigma'''_{\text{cut}}}\Bigg) 
\end{multline}
where 
\begin{align*}
    \sigma'''_{cut}&= \sigma''_{cut}\sqrt{1+ \frac{(a_{\text{cut}}+\gamma_x + \delta_x b_{\text{cut}})^2\text{Var}[\hat{x}_s]}{{\sigma''_{cut}}^2}}\\
    \gamma_x &= \frac{\text{Cov}[\hat{m}_s,\hat{x}_s]}{\text{Var}[\hat{x}_s]}\\
     \delta_x &= \frac{\text{Cov}[\hat{c}_s,\hat{x}_s]}{\text{Var}[\hat{x}_s]}  
\end{align*}

%%%%%%%%%%%%%%%%%%%%%%%%%%%%%%%%%%%%%%%%%%%%%%%%%%%%%%%%%%%%%%%%%%%%%%%%%%%%%%%
%%%%%%%%%%%%%%%%%%%%%%%%%%%%%%%%%%%%%%%%%%%%%%%%%%%%%%%%%%%%%%%%%%%%%%%%%%%%%%%
% DELETE THIS PART. DO NOT PLACE CONTENT AFTER THE REFERENCES!
%%%%%%%%%%%%%%%%%%%%%%%%%%%%%%%%%%%%%%%%%%%%%%%%%%%%%%%%%%%%%%%%%%%%%%%%%%%%%%%
%%%%%%%%%%%%%%%%%%%%%%%%%%%%%%%%%%%%%%%%%%%%%%%%%%%%%%%%%%%%%%%%%%%%%%%%%%%%%%%
%%%%%%%%%%%%%%%%%%%%%%%%%%%%%%%%%%%%%%%%%%%%%%%%%%%%%%%%%%%%%%%%%%%
%%%%%%%%%%%%%%%%%%%%%%%%%%%%%%%%%%%%%%%%%%%%%%%%%%%%%%%%%%%%%%%%%%%%%%%%%%%%%%%

\end{document}